\documentclass[conference]{IEEEtran}
\usepackage{hyperref}
\usepackage[table,xcdraw]{xcolor}
\usepackage{balance}
\usepackage{cite}

\ifCLASSINFOpdf
   \usepackage[pdftex]{graphicx}
\else
\fi
\begin{document}
\title{Beware the Normative Fallacy}

\author{
\IEEEauthorblockN{Christoph Becker}
\IEEEauthorblockA{ Faculty of Information\\
University of Toronto \\
Canada\\
christoph.becker@utoronto.ca}

% \IEEEauthorblockN{Christoph Becker}
% \IEEEauthorblockA{ Faculty of Information\\
% University of Toronto \\
% Canada\\
% Email:christoph.becker@utoronto.ca }
}

\maketitle

\begin{abstract}
Behavioral research can provide important insights for SE practices. But in performing it, many studies of SE are committing a normative fallacy – they misappropriate normative and prescriptive theories for descriptive purposes. 

The evidence from reviews of empirical studies of decision making in SE suggests that the normative fallacy may is common. This article draws on cognitive psychology and behavioral economics to explains this fallacy. Because data collection is framed by narrow and empirically invalid theories, flawed assumptions baked into those theories lead to misleading interpretations of observed behaviors and ultimately, to invalid conclusions and flawed recommendations. 

Researchers should be careful not to rely solely on engineering methods to explain what people do when they do engineering. Instead, insist that descriptive research be based on validated descriptive theories, listen carefully to skilled practitioners, and only rely on validated findings to prescribe what they should do.

\end{abstract}

%requirements engineering, critical system heuristics, system thinking 

% \fi
\IEEEpeerreviewmaketitle

%\section{Introduction}

\section{Introduction: Describing and Prescribing Behavior in SE}
The emerging focus on behavioral studies of software engineering is important: Without understanding what happens in engineering practice, we cannot hope to improve it. But which type of knowledge provides the foundation of such research? Put simply, engineering \textit{prescribes} what people should do – it develops normative frameworks such as methods that define what should be done and how. In contrast, studies of behavior \textit{describe} and explain what people do in practice. In behavioral studies of engineering, these two modes overlap, because they strive to describe and explain what happens in practice in order to prescribe what people could do better, and how, for some standard of evaluation. For example, many studies design and deploy new artifacts such as methods and tools into industrial contexts, then study how they impact performance. But practitioners frequently disregard newly developed methods introduced by academic research \cite{dittrich_what_2016}. Some teams may not use the artifacts, or use them differently than expected. How are the researchers going to interpret and explain this? Why did this team not adopt the method we developed? What did they do instead? How could we still help them to perform better in practice? 

To answer these questions, behavioral researchers collect data about professional practice. When they organize this empirical part of their study, they often rely on the toolbox of theories they used to design the methods. The theories we need to \textit{describe} what people do, however, are of fundamentally distinct nature and origin than the engineering methods that \textit{prescribe} what they should do. When it comes to how people make decisions, these theories even carry mutually incompatible assumptions. As a result, the tension between description and prescription can lead behavioral researchers to misunderstand practice in subtle but important ways.

The principal focus on description, explanation and analysis on one hand, and prescription or action on the other hand, can be used to distinguish two paradigms: \textit{Empirical} research aims to describe and explain behavior, while \textit{normative} research establishes standards to evaluate behavior \cite{ralph_two_2018}. Normative models such as SE methods, process models and quality models establish standards for evaluation. But there is a significant grey area because the object of empirical SE research is prescribed by normative models. For example, SE methods are sometimes treated as if they were ‘programs’ to be run by practitioners, even though they are more appropriately described as one of the resources that practitioners use in situated action \cite{dittrich_what_2016}.

Because research in SE often combines normative and empirical elements, even research that understands itself as empirical often relies on theories that are normative. As we will see, some of these theories have limited empirical validity –- or none. 

\section{How People Make Decisions}

The distinction between empirical and normative research, and between descriptive, explanatory and prescriptive theories, is not unique to SE. This article focuses on decision making because it is central to SE and as a field, has been bifurcated for decades as the tension between prescription and description split the field into separate schools  \cite{bell_decision_1989,beach_why_1993, klein_decision_1993,loewenstein_neuroeconomics_2008}. Behavioral SE has much to learn from that history.

Multi-criteria decision making (MCDM) methods are central to engineering methods. The normative framework of MCDM arose out of the mathematical theories of Bernoulli that prescribed how a theoretical agent \textit{should} take optimal choices between gambles under conditions of well-defined probabilistic uncertainty. The entire body of work in utility analysis is built on that starting point and has provided a foundation for countless methods in SE. MCDM methods prescribe how to analyze well-defined choice situations to identify which decision should be considered optimal, based on the assumption that the conditions and success criteria can be specified. According to this family of theories, an agent makes a decision by evaluating a set of options against a set of weighted criteria, uses this matrix to create a ranking of options, then selects the best option out of the set.

The underlying theory – now called ‘rationalistic’ in the field of Judgment and Decision Making –was never validated as a descriptive framework for human thought. In the words of Nobel prize winner Kahneman and Tversky, ``The modern theory of decision making under risk emerged from a logical analysis of games of chance rather than from a psychological analysis of risk and value. The theory was conceived as a normative model of an idealized decision maker, not as a description of the behavior of real people... [it] does not provide an adequate foundation for a descriptive theory of decision making'' \cite{tversky_rational_1986}.

On the contrary, ample evidence demonstrates that it is inconsistent with the reasoning processes of the human mind \cite{beach_why_1993}. Large-scale field studies in decision making observed and interviewed professionals in a wide range of fields, including military, surgery, firefighting, paramedics, and engineering. The studies collected overwhelming evidence showing that highly performing professionals did not evaluate multiple options against multiple criteria to compare them, rarely ranked options, and rarely selected an option from a set. Instead, they used their highly developed perceptual skills to match cues in the environment to patterns in their experience in order to generate one plausible course of action. They then used mental simulation to predict what would happen if they pursued it, and they adapted, adopted, or dropped one action at a time in sequence, until they found one that was satisfactory. In doing so, they often outperformed rationalistic approaches at a fraction of the time \cite{klein_sources_1998}. 

Rationalistic models are appealing to researchers, because their simple mathematical formulas promise a rigorous model of human behavior that supports the collection of data, the detection of deviations, and the design of interventions. As proponents of these models have emphasized, there is also normative value in these frameworks \cite{parnas_rational_1986}. But none of this provides descriptive validity \cite{beach_why_1993,shafer_savage_1986}.  In fact, rationalistic theories are descriptively inadequate in three important ways:

\begin{enumerate}

    \item Their \textbf{assumptions} –-  that preferences are fixed priors, that options are independent from each other, that preference relations are transitive, and that evaluation is independent of irrelevant alternatives –- are empirically invalid \cite{shafer_savage_1986,beach_why_1993,tversky_rational_1986}. The assumptions cannot be easily adjusted, because they are the theories’ foundational axioms \cite{shafer_savage_1986, beach_why_1993}.
    
    \item Their \textbf{predictions} are inconsistent with human behavior. In order to avoid dropping the entire body of theory, Tversky and Kahneman tuned the parameters of rational choice to create Prospect Theory  \cite{kahneman_prospect_1979}  and an influential research program on heuristics and biases. However, they retained the normative assumptions \cite{beach_why_1993}.
    
    \item The \textbf{methods} used throughout this period of behavioral research typically simplified choice situations into context-free vignettes and polled participants for their selection. It is a very efficient approach to data collection, but much of this award-winning work has no ecological validity --  it describes the behavior of ``people in the lab'', not the behavior of people ``in the wild'', and thereby distorts most of the factors that apply to real-world decision making \cite[36-50]{klein_decision_1993}.

\end{enumerate}

Despite their flaws, the appeal of rationalistic theories proved so strong that they underpinned most empirical research in cognitive psychology, behavioral economics and adjacent fields for decades. As a leading behavioral economist lamented, “50 years of dominance of the rational choice model has left so many important questions unanswered” \cite{loewenstein_neuroeconomics_2008}. SE emerged within that period, and its behavioral assumptions are firmly grounded in the normative camp. 

\section{The Normative Fallacy in SE}

In performing behavioral studies of SE, description and explanation need to be grounded not in a preconceived notion of what people are supposed to do but situated in their empirical reality. By staying attuned to the difference, we can minimize the chance of committing the \textbf{normative fallacy} \cite{campbell_normative_1970} in empirical research: the inappropriate use of normative or prescriptive theory for descriptive or explanatory purposes. 
 
Suppose we study the behavior of participants facing a risky software project situation. A series of 20 bugs were identified in short sequence in a new system under development. They appear somehow related, but it is unclear how. One is critical, 19 are severe. The team considers the business value of fixing one critical bug equivalent to fixing 4 severe bugs. Most team members want to focus all attention on the critical bug first, but one claims to know how to fix all 20. Our study participants must choose between two competing strategies: Strategy (a) has a $90\%$ chance of resolving the critical bug in the system, Strategy (b) has a $33\%$ chance of fixing all bugs. Many participants readily choose (a). According to the normative model, this is strictly speaking an ‘error', because in its calculation, the \textit{expected value} of option (b) is much higher, at  1.9 over 0.9. (The empirical validity of that calculation is dubious. Even statistically, the expected value is not a reasonable approximation of a single gamble, only of a long series of identical gambles \cite{shafer_savage_1986}.) 
 
The normative fallacy comes into play when we frame research questions and collect data. If we set out to \textit{describe} how participants make their choice, should we ask them ``how did you compute the value of each option?'' Doing so would mean committing the normative fallacy. Instead, we could ask an open-ended question: ``How did you reason about this?''. If given a chance, participants will rightfully introduce concerns outside the framing of the gamble that make it perfectly reasonable for them to prefer option (a). They might say that it can be communicated more effectively to the stakeholders; hope that fixing the critical bug produces insights that help fixing the others later; and believe that resolving the critical bug first will reduce the stress on the team. These judgments transcend the original framing of the gamble and situate it in a broader context. They seem much more reasonable then the normative assumptions of rational choice.  

In behavioral SE studies, however, the normative fallacy appears common. Systematic literature reviews in key areas suggest that many studies in SE use prescriptive theories to collect and interpret data about behavior for descriptive and explanatory purposes \cite{becker_intertemporal_2017,becker_trade-off_2018}.  In an extensive systematic literature review on Technical Debt decisions, we found that of the few studies that performed descriptive empirical work, all relied exclusively on rationalistic decision making theories \cite{becker_trade-off_2018}. For example, when a peer-reviewed empirical study in trade-off decision making asked participants about their use of reasoning, the questions framed the situation exclusively using the empirically invalid framework of rational choice:
 
 \begin{itemize}
     \item ``What factors are considered when you make a decision about when to fix a defect?''
  \item `'How are these factors weighted?''
 \end{itemize}
 
 These questions may appear perfectly normal, but only if we take the normative theories of rational choice as descriptively valid. They are not, so the questions above simply fail to adequately capture the participants’ reasoning \cite{beach_why_1993}. Examples of similarly structured questions seem common in SE. A second systematic review about studies of trade-off decisions in SE came to the same conclusion \cite{becker_intertemporal_2017}. Again, data collection in empirical studies was predicated on the narrowly framed concepts of rationalistic decision making: factors, weights, ranking, and choice. Broader considerations of reasoning, including the role of expertise, experience, cognition, incentives, mental simulation, judgment and perception were not considered. 
 
 This is not to say that participants in this example \textit{should not} consider a set of factors – simply that their individual and team reasoning processes will take forms that will remain invisible to those who perform the study \cite{klein_decision_1993}. Crucially however, because these questions are posed in the context of a scientific study by an academic research team with scientific credentials, the participants certainly provide answers to them. In doing so, they will retroactively construct plausible explanations. Papers uncritically reporting these answers, though not empirically valid, often pass peer review, because most reviewers in SE are not trained in behavioral research or psychology. The findings are then cited to support further normative research. Imagine if the study above had instead asked an open-ended question: ``How did you decide when to fix a defect?'' 
 
 In cognitive psychology, Klein’s studies provided a pivotal turning point toward naturalistic decision making. Klein recounts vividly how difficult it was for his research teams to drop the normative assumptions and listen to the data. It took remarkable insistence on the part of their study participants to push back against the assumptions carried by normative theories, especially the assumption that decision making functions by comparing alternatives \cite{klein_sources_1998}. 
 
 It is time that SE research follows this lead: When studying how software professionals act in industry settings, do not be led astray by the normative frameworks of software engineering \cite{dittrich_what_2016}. Instead, rely only on theories and methods with proven empirical validity. When studying decision making, emphasize the importance of naturalistic decision making theories, as at least one study has done \cite{zannier_model_2007}.
 
 \section{Listen to the Data}
 
 Just as in behavioral economics, the dominance of the rational choice model in software engineering has left many important questions unanswered. This is just as much an issue of practice as it is an issue of theory or methodology, because the normative fallacy carries an important corollary: Deviations from the normative model of decision making are generally labelled as ‘bias' and ‘error'. But that disregards the well-established fact that the normative model itself has no descriptive validity in the first place. 
 
 Our central example of the normative fallacy in SE is the reliance on rationalistic decision making theories in behavioral SE. Additional examples abound. The normative theory of process models is used widely to structure data collection of empirical studies, to interpret and analyze the data, and to describe what happened in these projects. But the theory is derived from normative origins and lacks empirical validity \cite{ralph_sensemaking-coevolution-implementation_2015}, so many of these studies may well be committing the normative fallacy. This would render their findings questionable at best and invalid at worst, and emphasizes the relevance of competing, empirically grounded theories \cite{ralph_two_2018}.
 
 Similar normative fallacies undermine studies of method adoption. The frequently observed disregard of practitioners for new methods \cite{dittrich_what_2016} may often be based on a well-founded judgment call grounded on expansive experience and a holistic assessment of a new method’s value. Yet, researchers focused on normative ideals may be quick to dismiss it as “gut feeling, rather than hard data gathered through proper measurement”, as one study writes. Cognitive psychology has long rehabilitated what this quote dismisses as ‘gut feeling’ as sophisticated forms of reasoning that transcend rationalistic frameworks, encompass them as one possible strategy among many, and apply them for the narrowly specified situations for which they are useful \cite{klein_sources_1998}. If researchers choose an empirically invalid theory as the baseline, we should locate error and bias firmly on their side. 
 
 \section{Avoid the Normative Fallacy}
 Currently, a large segment of behavioral studies of SE may be committing a normative fallacy: they misappropriate prescriptive theories for descriptive purposes. It is important to note that SE is not alone in incorrectly basing empirically oriented research about decision making on rationalistic theories with dubious merit -– other disciplines have grappled with this too.

The normative fallacy has far-reaching consequences: Because data collection is framed by narrow and empirically invalid theories, flawed assumptions baked into those theories lead to misleading interpretations of observed behaviors and ultimately, to invalid conclusions and flawed recommendations. As a side effect, the sound judgment of practitioners is sometimes dismissed as a defect to be fixed. But methods to support practice will be most effective when they are based on rigorous studies of practice using appropriate frameworks \cite{dittrich_what_2016}. 

Practitioners’ judgment and insight should be central to behavioral research in SE. Researchers should be careful not to rely solely on engineering methods to explain what people do when they do engineering. Instead, they must insist that descriptive research be based on validated descriptive theories, listen carefully to what skilled practitioners say and do, and only rely on validated findings to reason about prescribing what they should do.

Practitioners should similarly watch out for the normative fallacy and avoid acting on research that is not appropriately situated in real practice. When deliberating, keep in mind that a resistance to simplistic reduction is often not a sign of irrationality but a trait of sound judgment.

\balance
 
\bibliographystyle{IEEEtran}
\bibliography{NormativeFallacy}

\end{document}